\documentclass[fleqn,12pt,twoside]{article}
\usepackage[headings]{espcrc1}
\readRCS
$Id: espcrc1.tex,v 1.2 2004/02/24 11:22:11 spepping Exp $
\usepackage{graphicx}
\usepackage[figuresright]{rotating}

\usepackage{blopeps}
\blopps{1mm}
\bloplw{0.25mm}

\def\ET{\mbox{$E_T$}}
\def\pT{\mbox{$p_T$}}

\def\sqrtsNN{\mbox{$\sqrt{s_\mathrm{_{NN}}}$}}
\def\sqrts{\mbox{$\sqrt{s}$}}
\def\Npart{\mbox{$N_{part}$}}

\def\pizero{\mbox{$\pi^0$}}
\def\kzeros{\mbox{$K^0_s$}}

\def\vtwo{\mbox{$v_2$}}

\def\RAA{\mbox{$R_{AA}(\pT)$}}

\def\RCP{\mbox{$R_{CP}(\pT)$}}

\def\pTtrig{\mbox{$\pT^{trig}$}}
\def\pTassoc{\mbox{$\pT^{assoc}$}}

\def\qhat{\mbox{$\hat{q}$}}

\newcommand{\AmS}{{\protect\the\textfont2
  A\kern-.1667em\lower.5ex\hbox{M}\kern-.125emS}}

\def\lsim{\mathrel{\rlap{\lower4pt\hbox{\hskip1pt$\sim$}}
    \raise1pt\hbox{$<$}}}                
\def\gsim{\mathrel{\rlap{\lower4pt\hbox{\hskip1pt$\sim$}}
    \raise1pt\hbox{$>$}}}                
\title{High \pT\ in Nuclear Collisions at the SPS, RHIC, and LHC}

\author{P. Jacobs
\address{Lawrence Berkeley National Laboratory,\\
1 Cyclotron Road,\\
Berkeley, CA\\
USA 94720} and M. van Leeuwen \addressmark}

\begin{document}

\maketitle

\begin{abstract}

We review recent progress in the study of medium-induced modification
of jet fragmentation in high energy nuclear collisions at the SPS and
RHIC and present an outlook on jet physics at the LHC.

\end{abstract}

\section{Introduction}

Partonic energy loss is a potentially sensitive tomographic probe of
matter produced in high energy nuclear collisions, generating
observable modification of the fragmentation patterns of jets (``jet
quenching''). Measurements at RHIC have revealed large medium-induced
suppression at high transverse momentum (high \pT) of both the
inclusive hadron yields \cite{STAR_raa,PHENIX_raa} and of back-to-back
hadron pairs \cite{STAR_btob}. The principal energy loss mechanism
underlying these effects is commonly thought to be medium-induced
gluon bremsstrahlung, which is expected to dominate collisional
(elastic) energy loss for very energetic partons \cite{RadvsElastic}.

The effects of medium-induced radiation have been calculated in
various frameworks: multiple soft scattering (BDMPS \cite{BDMPS}), few
hard scatterings (GLV \cite{GLV}), twist expansion (Wang and Guo
\cite{GuoWang}), and light-cone path integral approach (Zakharov
\cite{Zakharov}).
In the case of multiple soft scattering the medium is
characterized by a single transport coefficient $\qhat=\mu^2/\lambda$,
where $\mu$ is the average momentum kick of a gluon interacting in the
medium and $\lambda$ is its mean free path. The gluon radiation
spectrum is suppressed relative to the Bethe-Heitler spectrum due to
coherence effects,
leading to medium-induced radiated energy
$\Delta{E}_{medium}\sim\alpha_{s}\qhat{L^2}$
\cite{BaierReview,SalgadoWiedemann}. Longitudinal expansion of the
medium reduces the length dependence to $\Delta{E}_{medium}\sim{L}$
while finite partonic energy truncates the medium-induced radiation
spectrum, resulting in an energy-dependent energy loss. Although the
induced radiation spectrum differs in detail between the BDMPS and GLV
approaches, the total energy loss is similar for comparable medium
properties \cite{SalgadoWiedemann}. For conditions relevant to RHIC
collisions the energy is lost in both cases to a moderate number
($\sim3$) of radiated gluons having moderate energy ($\sim0.1-1$ GeV)
\cite{SalgadoWiedemann}.

This picture is conceptually appealing and pQCD-based calculations
incorporating medium-induced bremsstrahlung reproduce much of the
published data on high \pT\ hadron production in nuclear
collisions. Nevertheless, it is important to ask to what extent the
data {\it require} this description to be the correct one. Are its
detailed predictions, such as the expected $L^2$
dependence of $\Delta{E}$, observed? Alternatively, does {\it collisional}
energy loss play a significant role in the finite kinematic regime of
RHIC \cite{ThomaMustapha}?

We review recent measurements of partonic energy loss in
hot matter, emphasizing results shown at this conference, and compare
their systematic behavior to theoretical expectations. The variables
at our disposal are energy, centrality and \pT\ dependence of
high-\pT\ hadron production and correlations. We concentrate on the
highest available \pT\ at mid-rapidity.

\section{SPS}

\begin{figure}
\includegraphics[width=0.5\textwidth]{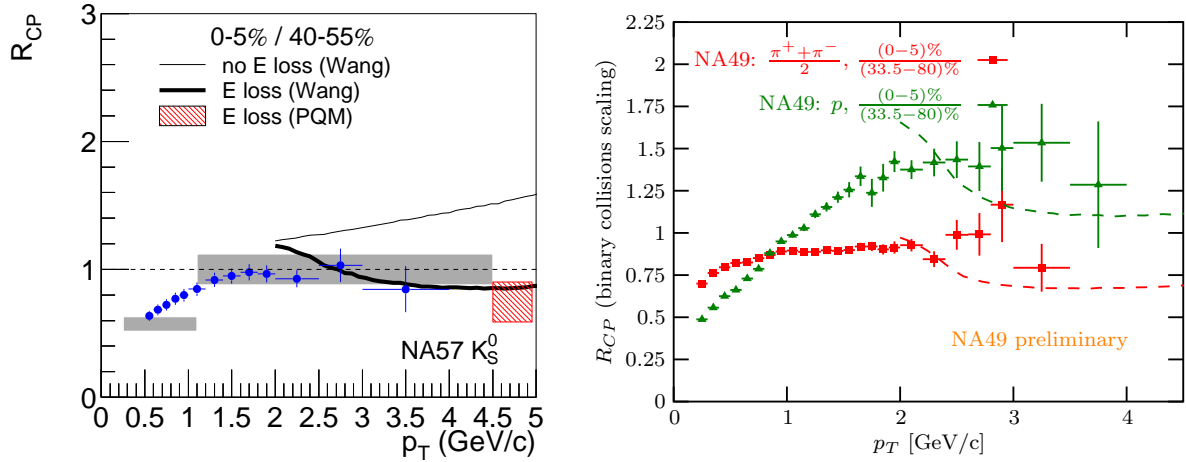} 
\scriptsize\blopeps[width=8cm, height=6cm]{figs/NA49_RCP.beps}

\caption{Nuclear modification factor \RCP\ measured at the
  SPS and compared to model calculations \cite{WangSPS}. Left panel: \kzeros\ from NA57
  \cite{NA57}; Right panel: charged pions and protons from NA49
  \cite{NA49}.}
\label{fig:RAASPS}
\end{figure}

A long-standing puzzle concerns the magnitude of partonic energy loss
in nuclear collisions at the SPS. WA98 data \cite{WA98} indicated a
large enhancement of inclusive \pizero\ production at $\pT\sim3-4$
GeV/c in central Pb+Pb relative to p+p collisions when normalized
per binary collision (\RAA), although a suppression was observed in
central relative to peripheral nuclear collisions (\RCP, also
normalized per binary collision) by the same experiment.

The p+p spectrum used as a reference for \RAA\ at the SPS was not
directly measured but was based on an extrapolation from measurements
at higher \sqrts. Some of the datasets used in this extrapolation are
only marginally consistent, leading to large uncertainties in the p+p
reference spectrum. A recent re-evaluation of the extrapolation by
d'Enterria \cite{dEnterriaSPS} found a reduced \RAA, more
consistent with \RCP. The medium density inferred from the new \RAA\
values is now also consistent with expectations from Bjorken energy density
estimates \cite{VitevSPS,WangSPS}.

New measurements of high-\pT\ hadron suppression at the SPS were
presented at this conference. Fig. \ref{fig:RAASPS} shows \RCP\ for
\kzeros\ from NA57 \cite{NA57} (left panel) and charged pions and
protons from NA49 \cite{NA49} (right panel) compared to radiative energy loss
calculations. A marked Cronin effect is expected (``no E loss'' on
left panel) due to the steep \pT\ spectrum. Introduction of energy
loss in a medium, with gluon density scaling as $dN_{ch}/d\eta$
\cite{WangSPS}, results in good agreement between calculation and data.

These new data solve the high-\pT\ suppression puzzle at the SPS: the
medium densities inferred from bulk multiplicity and high-\pT\
inclusive hadron measurements are consistent. However, 
significant theoretical uncertainties remain due to a potentially large
Cronin effect at these lower energies.

\section{RHIC: inclusive yields}

Rich and initially unexpected phenomenology has emerged in the
``intermediate \pT'' region ($\pT\sim2-5$ GeV/c) of nuclear collisions
at RHIC. The enhancement of baryon relative to meson yields
\cite{OlgaSTAR,PHENIXBaryonMeson}, the approximate scaling of elliptic
flow \vtwo\ with the number of constituent quarks \cite{SorensenQM05},
and correlation measurements in this region \cite{AjitanandQM05}
suggest an interplay between hadronization of hard partons and the
bulk medium, as discussed extensively elsewhere at the conference
\cite{SorensenQM05,HwaQM05}. Here we will concentrate on the region
$\pT \gsim 6$~GeV, i.e. above the region of the anomalous enhancement
of the baryon/meson ratio \cite{OlgaSTAR,PHENIXBaryonMeson}, to avoid
these complex hadronization phenomena that may obscure the
modification of fragmentation due to partonic energy loss.


\begin{figure}
\begin{minipage}{0.45\textwidth}
\vspace{0.4cm}
\includegraphics[width=\textwidth]{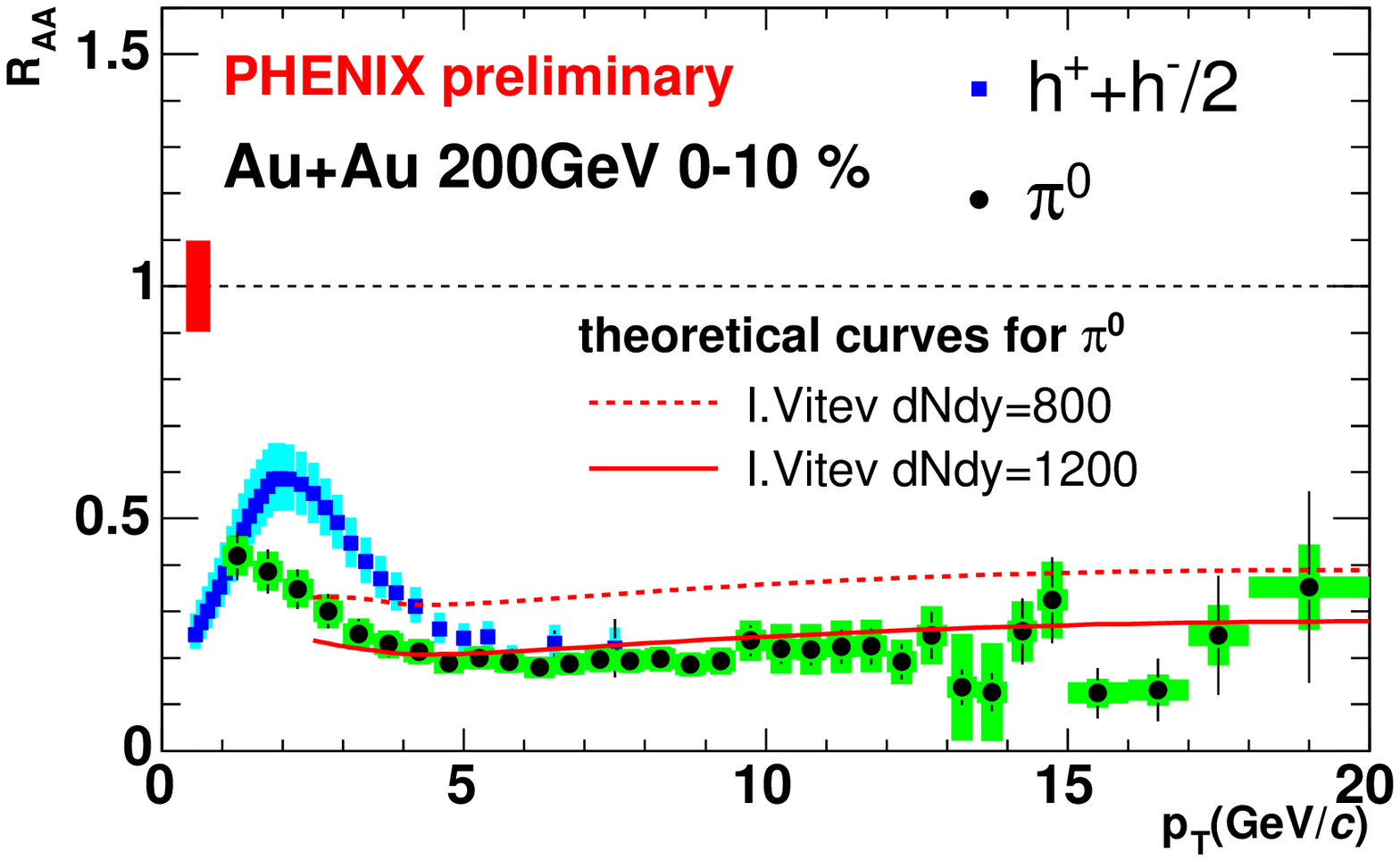}
\end{minipage}
\hfill
\begin{minipage}{0.5\textwidth}
\includegraphics[width=\textwidth]{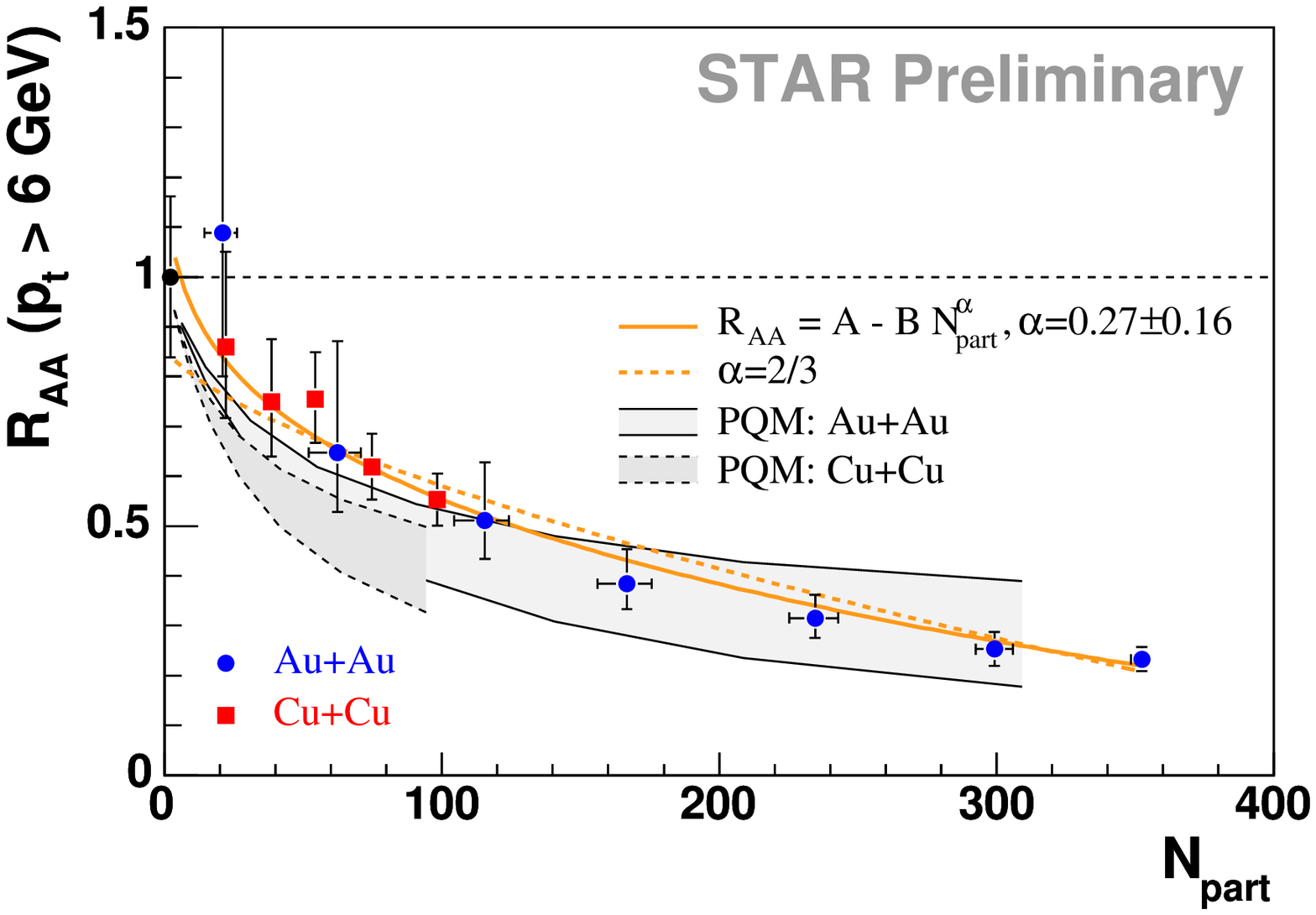}
\end{minipage}
\caption{Inclusive high \pT\ hadron suppression at RHIC. 
Left: \RAA\ for \pizero\ and charged hadrons in central Au+Au
collisions from PHENIX \cite{PHENIXRAA}. The curves show suppression
calculated in the single hard scattering approach \cite{VitevSPS}.
Right: \Npart\ dependence of \RAA\ at $\pT>6$ GeV for charged hadrons
from Cu+Cu and Au+Au collisions measured by STAR \cite{DunlopQM05},
compared to calculations in the multiple soft scattering approach \cite{DaineseLoizides}.
}
\label{fig:HadronSuppression}
\end{figure}


Fig. \ref{fig:HadronSuppression}, left panel, shows the most recent
measurement of \pizero\ \RAA\ from the high-statistics 2004 Au+Au run
at RHIC \cite{PHENIXRAA}. The kinematic reach now extends to
$\pT\sim20$ GeV/c, well beyond the intermediate \pT\ region. The
observed
\pT-independence of the suppression over this broad range is well
described by partonic energy loss models in both the few hard
scattering \cite{VitevSPS} (dashed and solid line) and multiple soft
scattering \cite{Eskola05,SalgadoQM05} (not shown) limits, requiring
medium densities $dN_g/dy\sim1000-1200$ and $\qhat\sim5-15$ GeV$^2$/fm
respectively. The transport coefficient \qhat\ and gluon density
$dN_g/dy$ in these models are time-averaged quantities. The connection
between these averaged quantities and the gluon or energy density of the medium at early
time is discussed elsewhere \cite{Eskola05,SalgadoQM05}.

The path length dependence of hadron suppression is a key test of its
underlying mechanism. During the 2005 RHIC run, large samples of Cu+Cu
collisions at \sqrtsNN=200 GeV were collected to measure high-\pT\
particle production in this smaller system with similar statistical
reach as the long 2004 Au+Au run. First analyses of a subset of the
total Cu+Cu dataset were presented at this
conference. Fig~\ref{fig:HadronSuppression}, right panel, shows the
centrality dependence of \RAA\ for $\pT>6$ GeV charged hadrons, from
Cu+Cu collisions (squares), compared to existing results from Au+Au
collisions (circles) \cite{DunlopQM05}. The most significant
comparison is to central Cu+Cu, where the Glauber calculation
uncertainties are smaller than for the semi-peripheral Au+Au collision
data with similar $\Npart\sim100$. The data indicate that the
inclusive hadron suppression is a function solely of \Npart, with no
observable sensitivity to the shape of the reaction zone. The lines
indicate phenomenological fits to characterize the \Npart\ dependence
of \RAA. The data prefer a decrease with $\Npart^{1/3}$, although the
more commonly expected $\Npart^{2/3}$ scaling is not strongly
excluded. In general, the observed scaling behavior results from the
combined effects of the spectrum shape, the collision geometry, and
the path-length dependent energy-loss distribution. The gray bands
indicate the results of a full calculation incorporating these effects
\cite{DaineseLoizides}, which reproduces the common suppression in
Cu+Cu and Au+Au at the same \Npart\ but gives slightly larger
suppression at low \Npart\ than observed in the data.

\section{RHIC: correlations}

Sensitivity of inclusive hadron suppression measurements to the
properties of the medium is limited due to a surface bias: for a dense
system, the observed hadrons are preferentially the fragments of
partons produced near the surface and headed outwards, which 
suffer less than average energy loss\cite{BaierQM02,Eskola05}. The
observed $\RAA\sim0.2$ in central Au+Au collisions is reproduced by
calculations with a broad range of $\qhat\sim5-15$ GeV$^2$/fm
\cite{Eskola05}, suggesting that its value is essentially geometric in
origin. A more detailed view of the medium is obtained from
back-to-back correlations of hadron pairs, where the surface bias can
be reduced by increasing the \pT-threshold for associated particles on
the away-side. This leads to counterbalancing biases from the
requirements on the trigger and recoil, providing a new and more
sensitive probe of the medium.

It has been known for some time that the back-to-back high \pT\
di-hadron yield is strongly suppressed in the most central Au+Au
collisions \cite{STAR_btob}. However, the kinematic cuts of this first
measurement ($\pTtrig>4$ GeV/c for the trigger and $\pTassoc>2$ GeV/c
for the associated hadrons) were relatively low, giving rise to a
large combinatorial background. Due to the large background and the
strong suppression, a differential measurement of the recoil yield was
not possible and only an upper limit could be established. Lowering
\pTassoc\ reveals an excess and broadening of the recoil yield
\cite{STARCorrLowpTtoB}, to the extent that the correlation structure
is compatible with expectations from simple momentum conservation
without additional dynamic correlations. For these lower \pTassoc,
however, the backgrounds are yet larger and the uncertainties in the
background yield and the flow modulation of the background make a
quantitative measurement of the excess difficult. Qualitatively, these
observation are compatible with a picture in which strong partonic
energy loss in the core of the reaction volume softens the
fragmentation. The
response of the medium to the energy loss then becomes a key issue,
which was discussed extensively elsewhere at this conference
\cite{AjitanandQM05,UleryQM05,GrauQM05}.

\begin{figure}
\centering
\includegraphics[width=0.5\textwidth]{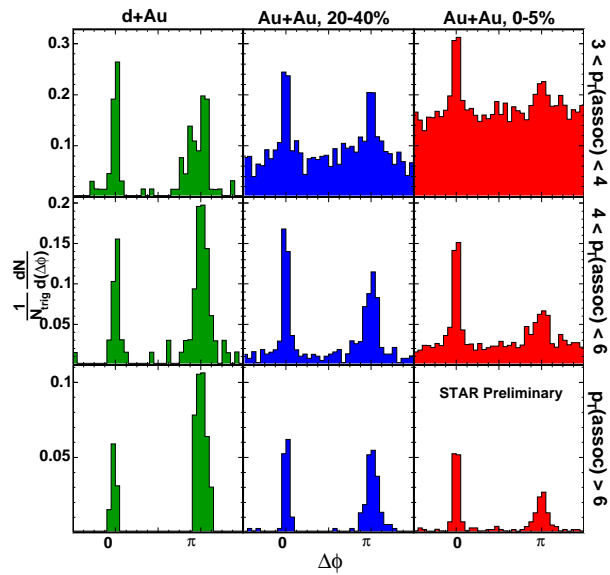}
\caption{Per-trigger correlated yield for d+Au and Au+Au collisions in
  two different centrality ranges, with $\pTtrig>8$~GeV. The different
  rows show the correlation for different ranges in \pT\ of the
  associated particle \cite{MagestroQM05}.}
\label{fig:STARCorr}
\end{figure}

New analyses from the long 200 GeV Au+Au run at RHIC now provide much greater
\pT\ reach for correlation studies, for the first time extending
beyond the intermediate \pT\ region. Fig. \ref{fig:STARCorr} shows
azimuthal correlations of charged hadrons for $\pTtrig>8$
GeV/c and varying threshold on \pTassoc\
\cite{MagestroQM05}. It is seen from the figure that the combinatorial
background is negligible for the highest cuts and a clear back-to-back
correlation signal emerges, for the first time enabling a quantitative
{\it differential} measurement of partonic energy loss.

\begin{figure}
\centering
\includegraphics[width=0.68\textwidth]{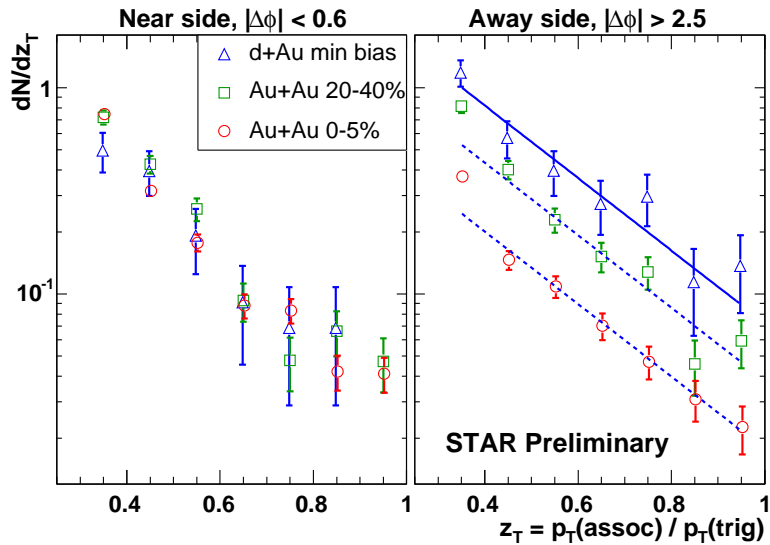}
\caption{Di-hadron fragmentation distributions on the near- (left
  panel) and away-side (right panel) for $8<\pTtrig<15$ GeV
  \cite{MagestroQM05} in d+Au collisions and Au+Au collisions at two
  different centralities. The solid line shows an exponential fit
  to the d+Au data. The dashed lines have the same slope, but are
  scaled down to match the Au+Au data. }
\label{fig:STARCorrDetails}
\end{figure}

This is further explored in Fig.~\ref{fig:STARCorrDetails}
\cite{MagestroQM05}, which shows distributions of the near- (left
panel) and away-side (right panel) associated hadrons as a function of
di-hadron fragmentation variable $z_T=\pTassoc/\pTtrig$ \cite{Wang04},
for $8<\pTtrig<15$~GeV trigger hadrons in d+Au and semi-central and
central Au+Au collisions. The lines in the right-hand panel indicate
an exponential fit to the d+Au data (solid line) which is scaled by
0.54 and 0.25 to match the semi-central and central Au+Au data
respectively (dashed lines).

On the near side, no significant variation of the yield or the
fragmentation distribution is seen between the systems. Calculations
in Ref. \cite{Majumder04} predict a strong enhancement of the
associated yield in central collisions due to large energy loss and
corresponding trigger bias effects, which is not observed in the data.

In contrast to the centrality invariance of the near-side correlation,
a strong suppression of the away-side yield is found in central collisions
at a level ($\sim25$\%) that is numerically similar to \RAA. This
away-side suppression is however not accompanied by observable
modification to either the longitudinal fragmentation distribution
(lines in Fig.~\ref{fig:STARCorrDetails}) or the azimuthal distribution (see
Fig. \ref{fig:STARCorr} and \cite{MagestroQM05}).

Suppression that is independent of $z_T$ for $z_T>0.4$ is in qualitative
agreement with calculations by Wang \cite{Wang04}, though
quantitatively the measured suppression is stronger than
predicted. Strong suppression could be accompanied by broadening of
the recoil distribution in the event-averaged two particle correlation
function, due either to large-angle emission of fragments from
radiated gluons or to medium-induced acoplanarity of the dijets. In
the multiple soft scattering approach the relation between energy
loss $dE/dx$ and induced acoplanarity has
the general form \cite{BaierpTBroadening}

\begin{equation}
-\frac{dE}{dx}=\frac{{\alpha_s}{N_c}}{8}\cdot\langle\pT^2\rangle_{jet},
\end{equation}

\noindent
where $\langle\pT^2\rangle_{jet}$ refers to the momentum transverse to
the initial direction of the hard parton that is acquired from
interactions in the medium. Detailed calculations of the broadening
using this approach are not yet available. A calculation for
large energy loss in the GLV framework \cite{Vitev05} predicts that
induced radiation will dominate the recoil hadron distribution up to
high $\pT\sim10$ GeV/c, leading to strong azimuthal broadening. In
contrast, Fig.~\ref{fig:STARCorr} shows no significant angular
broadening in this kinematic range (see also \cite{MagestroQM05}).

Current model calculations invoke independent gluon emission to
calculate integrated energy loss \cite{SalgadoQM05} and may not
incorporate both radiative and elastic energy loss in a consistent
way. An improved theoretical framework is likely required to calculate
accurately the induced acoplanarity. On the experimental
side, higher-order multi-particle correlations
\cite{AjitanandQM05,UleryQM05} may be able to distinguish broadening due to
additional radiation and induced acoplanarity of the jets, which is
not possible with event-averaged two-particle correlations.

Significant suppression of the recoil yield without measurable changes
in the longitudinal and azimuthal correlations would arise if this
measurement is sensitive primarily to hadronic fragments of partons
that had little or no interaction with the medium. This bias arises
naturally in radiative energy loss calculations, which predict finite
probability to emit {\it zero} gluons in a medium of finite length
\cite{SalgadoWiedemann}, with emission of a moderate energy gluon
leading to sizeable suppression in the relatively high \pT\ region of
this analysis. A quantitative evaluation of these effects, including a
realistic nuclear geometry and path length dependent energy loss
distributions, is given in Ref.~\cite{DaineseLoizides}, mainly in the
context of single-hadron suppression. These calculations indicate
that the probability for both the trigger and recoil jet to have
little interaction with the medium is small, and a contribution is
expected from partons that experience significant interaction with the
medium. Refinement of these and other calculations within the
constraints of the new data presented at this conference will help
elucidate the mechanisms underlying jet quenching and the properties
of the medium. A tantalizing prospect from these jet quenching studies
is an experimentally measured upper bound on \qhat, which could
constrain the number of underlying degrees of freedom of the medium
\cite{RajaMuller}.

\section{Jet Physics at the LHC}

\begin{figure}
\centering
\includegraphics[height=0.44\textheight,clip=]{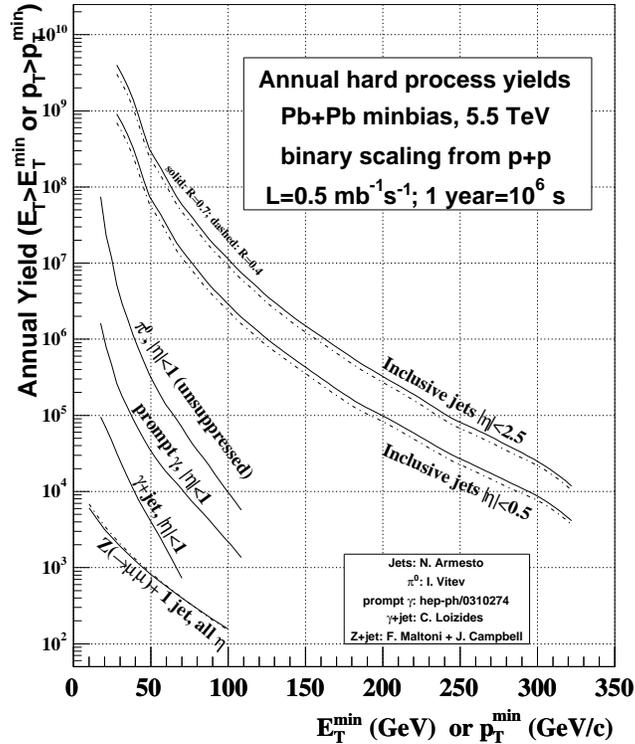}
\caption{Hard process rates at the LHC \cite{LHCrates,YellowReporthotons}. Solid and dashed lines for inclusive jets indicate 
rates vs \ET\ integrated over phase space cone of radius R=0.4  and 0.7 .}
\label{fig:LHCRates}
\end{figure}

The CERN LHC is currently scheduled to commission p+p collisions in
2007, with the first Pb+Pb run at \sqrtsNN=5.5 TeV in 2008
\cite{HansAkeQM05}. The factor 30 increase in collision energy
relative to RHIC generates a huge increase in kinematic and
statistical reach for hard probes. Figure~\ref{fig:LHCRates} shows the
yield for various observables relevant to jet quenching studies
\cite{LHCrates,YellowReporthotons} expected from one LHC year of Pb+Pb
running ($10^6$ seconds) at nominal luminosity. Simple binary
collision scaling ($\propto A^2$) from calculated p+p cross-sections has been
applied, with no nuclear effects taken into account.

There will be statistically robust yields for jets well above $\ET\sim200$
GeV, providing logarithmically large energy variation over which to
study jet quenching effects. The huge statistics will enable the study of
rare, perturbatively calculable fragmentation channels such as very
hard hadron pairs with small angular separation, whose distributions
may be modified by medium effects. The large yield of high energy
jets raises the possibility that multi-particle or calorimetric
(quasi-)full jet reconstruction can be used even in the presence of
large backgrounds in central nuclear collisions. This potentially
recovers the energy radiated in gluons, allowing relatively unbiased
reconstruction of the jet energy and enabling complete
characterization of jet quenching without the complications of strong
trigger and geometric biases that are present in leading hadron and
di-hadron measurements as performed at RHIC.

Jet reconstruction with good energy resolution is not straightforward
in central nuclear collisions even at the LHC, however. For jets with
$\ET\sim50-100$ GeV, $\sim80\%$ of the charged track energy is
contained in a cone of phase space radius
$R=\sqrt{\delta\eta^2+\delta\phi^2}\sim0.2$ \cite{CDFJet}, while an
area of this size in a central Pb+Pb collision at 5.5 TeV may contain
$\sim75$ GeV \cite{EskolaLHCestimates} of uncorrelated energy from soft
particle production. Simulation studies of jet reconstruction using
unmodified fragmentation show that relatively small cone radii
$R\sim0.3-0.4$ provide optimum energy resolution \cite{ALICE_QM04},
though the models of both jet signal (PYTHIA) and background (HIJING)
in these studies do not capture the physics of jet softening and
broadening due to medium effects. Full understanding of jet
reconstruction capabilities and optimization of jet observables at the
LHC must await data.

A recent calculation incorporating medium effects in the MLLA parton
shower framework \cite{BorghiniMLLA} predicts an enhancement in
multiplicity due to jet quenching up to hadron $\pT\sim5-6$ GeV/c for
jets with $\ET\sim100-200$ GeV. This excess should be
measurable above background, enabling a detailed characterization of
the modification of the fragmentation function. Measurement of high
energy b-tagged and c-tagged jets can be contrasted to light hadron-led
jets, which arise dominantly from gluons, to exploit the different
color charge coupling of gluons and quarks to the medium (factor
9/4). A broad range of multi-hadron correlations will be accessible
which interpolate between leading hadron studies and full jet
reconstruction.

The golden channel for jet quenching is the coincidence measurement of
a jet recoiling from a gauge boson ($\gamma$ or Z). The boson does not
interact with the medium and therefore provides a clean calibration of
the momentum transfer in the interaction \cite{WangGammaJet}, enabling
measurement of the true fragmentation of the recoiling jet. While this
strategy remains attractive, the measurements are
challenging. Fig.~\ref{fig:LHCRates} shows that $\gamma$+jet rate is
statistically robust in Pb+Pb only for $\pT < 40$ GeV/c.  The
$\gamma$/\pizero\ ratio (the key experimental parameter) exceeds 10\%
only for $\pT > 50$ GeV/c in p+p, though \pizero\ suppression may
lower that bound to 20 GeV/c in central Pb+Pb
\cite{YellowReporthotons}. Most significantly, QCD fragmentation
photons may dominate the prompt photon yield up to 50 GeV or higher
\cite{ArleoLHCGammaJet}, complicating the interpretation of the photon
as a non-interacting messenger from the hard vertex. The Z+jet channel
is background-free but suffers from small cross section (see
Fig.~\ref{fig:LHCRates}) and will be statistically marginal at nominal
Pb+Pb luminosity.

\section{Summary}

The SPS and RHIC data presented at this conference significantly
extend previous jet quenching studies, in some cases providing
qualitatively new insights. The data are in broad agreement with
expectations based on the dominant paradigm of radiative energy loss,
but experimental tests of this correspondence are not yet definitive. 

A characteristic feature of radiative energy loss is the quadratic
path length dependence that arises from coherence effects. The new
Cu+Cu data, combined with existing Au+Au measurements, will provide
the most precise tests of this scaling. First hadron suppression
measurements in Cu+Cu are in rough agreement with radiative energy
loss models, but theoretical uncertainties now dominate this
comparison and further progress will come from the theory side.

New di-hadron correlation studies reveal a well-defined recoil peak at
high \pT{}, enabling the first differential measurements of jet
suppression in the medium. Many existing calculations miss essential
features of these data, perhaps due to approximations in the treatment
of nuclear geometry or energy loss. Comparison to the most complete
calculation available \cite{LoizidesQM05} gives a new constraint
$\qhat\sim5-7$ GeV$^2$/fm (using the non-reweighted version of this
calculation), though the uncertainty on this number is difficult to
estimate. Heavy flavor suppression, measured via non-photonic
electrons, may be larger than expected from radiative energy loss
alone \cite{charm_suppr}, raising the question of significant
collisional (elastic) energy loss.

The experimental study of jet quenching has reached a new level of
detail and precision, and interpretation of the striking effects that
have been observed is currently limited by theoretical
uncertainties. Heavy ion collisions at the LHC will open up a huge new
kinematic regime for jet quenching studies, with qualitatively new
observables available. While the LHC regime may provide better grounds
for quantitative theoretical predictions, a complete picture of jet
interactions with dense QCD matter must describe all of the phenomena
we observe, both at RHIC and at the LHC.



\end{document}